\documentclass[conference]{IEEEtran}
\IEEEoverridecommandlockouts
\usepackage{cite}
\usepackage{amsmath,amssymb,amsfonts}
\usepackage{multirow, multicol}
\usepackage{algorithmic}
\usepackage{graphicx}
\usepackage{siunitx}
\usepackage{textcomp}
\usepackage{xcolor}
\usepackage{hyperref}
\def\BibTeX{{\rm B\kern-.05em{\sc i\kern-.025em b}\kern-.08em
    T\kern-.1667em\lower.7ex\hbox{E}\kern-.125emX}}

\def\new[#1]{\textcolor{black}{#1}}

\begin{document}

\title{Gradient-based Optimisation of Modulation Effects}

\author{
\IEEEauthorblockN{Alistair Carson, Alec Wright}
\IEEEauthorblockA{\textit{Acoustics and Audio Group} \\
\textit{University of Edinburgh},
Edinburgh, UK \\
\href{mailto:alistair.carson@ed.ac.uk}{alistair.carson@ed.ac.uk},\href{mailto:alec.wright@ed.ac.uk}{ alec.wright@ed.ac.uk}}

\and

\IEEEauthorblockN{Stefan Bilbao}
\IEEEauthorblockA{\textit{STMS lab} \\
\textit{IRCAM-CNRS-Sorbonne University},
Paris, France \\
\href{mailto:bilbao@ircam.fr}{bilbao@ircam.fr}}
}

\maketitle
%Abstract
\begin{abstract}
Modulation effects such as phasers, flangers and chorus effects are heavily used in conjunction with the electric guitar. Machine learning based emulation of analog modulation units has been investigated in recent years, but most methods have either been limited to one class of effect or suffer from a high computational cost or latency compared to canonical digital implementations. Here, we build on previous work and present a framework for modelling flanger, chorus and phaser effects based on differentiable digital signal processing. The model is trained in the time-frequency domain, but at inference operates in the time-domain, requiring zero latency. We investigate the challenges associated with gradient-based optimisation of such effects, and show that low-frequency weighting of loss functions avoids convergence to local minima when learning delay times. We show that when trained against analog effects units, sound output from the model is in some cases perceptually indistinguishable from the reference, but challenges still remain for effects with long delay times and feedback. 
\end{abstract}

\section{Introduction}
Modulation effects including phasing, flanging, chorus and vibrato are widespread in music production and performance \cite{bode1984history}. These effects involve phase modulation of an audio signal controlled by one or more low-frequency oscillators (LFOs) \cite{DutilleuxDAFx}. 
Phasers typically use a cascade of first-order analog or digital all-pass filters and mix the dry and phase-modulated signals to produce a time-varying filter with a small number of peaks and notches in the spectrum \cite{JuliusOSmith1982}. The delays in flanger, chorus and vibrato effects are generally longer---on the order of several milliseconds---and implemented with interpolated digital delay lines \cite{JuliusOSmith1982}, analog bucket-brigade devices \cite{Sangter1969_BBD} or magnetic tape manipulations \cite{bartlett1970a, PASPWEB2010}. 

The design and optimisation of digital modulation effects to emulate a reference analog effect unit, such as a guitar pedal, is an example of virtual analog modelling \cite{DAFXVAchapter}. Such electronic circuit-based simulations can be accurate and lightweight \cite{Eichas2014, Huovilainen_2005, Giampiccolo2024_phase90, Holters2018_BBD} but are device specific. Neural network black-box models \cite{Ramirez2020_blackbox} offer more generality but at greater computational expense. Lighter-weight neural models have been presented in the form of recurrent neural networks conditioned on the LFO signal \cite{Wright2021}, which must be either manually measured \cite{Wright2021} or whose behaviour must be predicted using another neural network \cite{Mitcheltree2023}. In both cases, the models struggled to accurately implement the longer delay times in flanger effects \cite{Wright2021, Mitcheltree2023}. Recurrent neural networks have a strong inductive bias towards distortion and non-linear time-invariant effects \cite{Wright2020}, but their application to modulation effects---that are most often classified as linear time-varying \cite{DutilleuxDAFx}---is perhaps less suitable. 

Differentiable digital signal processing (DDSP) \cite{Engel2020, hayes_review_2023} offers a potential middle ground that exploits the benefits of automatic differentiation and well-known priors. Previous work \cite{Carson2023} presented a method for optimizing a DDSP phaser topology to model an analog reference device, including a differentiable LFO learned during training \cite{Carson2023}. Lee et al. \cite{Lee2024} extended this method to a general time-varying filter and showed that this could emulate both phaser and flanger effects. Both methods \cite{Carson2023, Lee2024}, however, operate in the time-frequency domain which imposes a minimum latency at inference. The DDSP phaser model \cite{Carson2023} was later extended to a time-domain implementation \cite{ycy2024diffapf}, but issues with numerical stability made training unpredictable. Other recent related work used a DDSP architecture to extract modulation signals from a target synthesiser sound \cite{Mitcheltree2025}.

Here we build on the DDSP phaser model \cite{Carson2023} and propose a novel extension to chorus and flanger effects. The model is trained using a time-frequency domain approximation of a time-varying linear filter, but unlike related works \cite{Carson2023, Lee2024}, at inference the model is implemented in the time-domain, requiring zero latency. An analysis of the challenges of optimising phase shifting effects is provided in Sec.~\ref{sec:grads}, and the proposed method is presented in full in Sec.~\ref{sec:model}. Experimental evaluation is provided for a simplified problem in Sec.~\ref{sec:toy}, and for virtual analog modelling in Sec.~\ref{sec:bf-2}, Sec.~\ref{sec:sv1} and Sec.~\ref{sec:smallstone}. The results from a perceptual evaluation are given in Sec.~\ref{sec:mushra}, and concluding remarks are provided in Sec.~\ref{sec:conclusion}. Python code, model weights and audio examples are provided online \footnote{\href{https://a-carson.github.io/modulation_fx/}{https://a-carson.github.io/modulation\_fx/}}.

\section{Phase estimation with gradient descent}\label{sec:grads}

\subsection{Estimating delay times}\label{sec:grad:delay}
The accurate and robust prediction of time-varying phase shifts is central to the problem of modelling flanger, phaser and chorus effects. Let us first consider a target system that applies a delay of $D \in \mathbb{R}^{+}$ samples (not necessarily integer) to a time series $x[n]$. This output is given by the convolution:
\begin{equation}\label{eq:delay_td}
    y[n] = \sum_{\tau=0}^{N-1 } \textrm{sinc}(\tau - D)x[n - \tau], \quad n = 0, \dots, N-1
\end{equation}
where $\textrm{sinc}(\eta) = \sin(\pi \eta)/(\pi \eta)$. For integer $D \leq N - N'$ where $N'$ is the index of the last non-zero sample in $x[n]$, expression \eqref{eq:delay_td} reduces to $y[n] = x[n-D]$. If the delay exceeds this, the output signal will be truncated, so sufficient zero padding is essential to avoid this. For non-integer $D$, the sinc function provides ideal band-limited approximation to the fractional delay \cite{Laakso1996}. In this case the impulse response has infinite-support and truncation is inevitable, but its effects can be lessend by increasing $N$ or using tapered window functions \cite{Laakso1996}.

In the frequency domain, the system \eqref{eq:delay_td} can be approximately expressed as:
\begin{equation}\label{eq:dft_delay}
Y(k) = 
\begin{cases}
     X(k)e^{-j 2 \pi k D / N}, \quad &0 \leq k \leq N/2 \\
      X(k)e^{j 2 \pi (N-k) D / N}, \quad &N/2 < k < N
\end{cases}
\end{equation}
where $X(k)$ is the $N$-length (for even $N$) discrete Fourier transform (DFT) of $x[n]$. The time-domain output can then be taken as the inverse DFT of $Y(k)$, but sufficient zero-padding is again essential here: for $N \geq D + N'$ and integer $D$, the systems \eqref{eq:delay_td} and \eqref{eq:dft_delay} are equivalent, but otherwise time-aliasing will occur.

\begin{figure}[t]
    \centering
    \includegraphics[width=\linewidth]{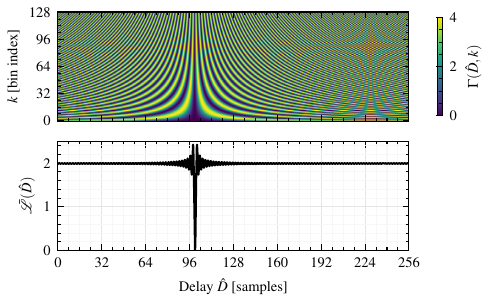}
    \caption{Loss surface $\Gamma(\hat{D}, k)$ (top) and its mean over bin index $\mathcal{L}(\hat{D})$ for a spectrally flat input signal (bottom) for DFT domain estimation of a delay of $D=100$ samples.}
    \label{fig:dft_delay_colormap}
\end{figure}

In this work we explore estimating the delay $D$ through gradient descent optimisation:
\begin{eqnarray}
    \quad \quad \quad \quad D = \textrm{arg}\min_{\hat{D}} \mathcal{L}(\hat{D})
\end{eqnarray}
where $\hat{D}$ is the optimization variable (the estimated value of the delay) and $\mathcal{L}(\hat{D})$ is the objective or loss function. The loss function can be defined either in the time-domain (TD) or frequency-domain (FD):
\begin{subequations}\label{eq:dft_delay_loss}
\begin{align}
    \mathcal{L}_{\textrm{TD}} (\hat{D}) &= \sum_{n=0}^{N-1} |y[n] - \hat{y}[n]|^2 \\
    \mathcal{L}_{\textrm{FD}} (\hat{D}) &= \sum_{k=0}^{N-1} |Y(k) - \hat{Y}(k)|^2
    % \mathcal{L}_{\textrm{FD}}(\hat{D})  &= \sum_{k=0}^{N/2}  |X(k)|^2 \left(1 - \cos\left(2\pi k\frac{D - \hat{D}}{N}\right)\right).
\end{align}
\end{subequations}
where $\hat{y}[n]$ is the input signal delayed by $\hat{D}$ samples and $\hat{Y}$ is its spectrum. As $N \to \infty$ (or integer $D \leq N - N'$) the above expressions are equivalent due to Parseval’s theorem \cite{oppenheim189_dsp}. Here we focus on FD optimisation, which has been used in other DDSP applications \cite{DalSanto2025_flamo, Lee2022}.
Taking only the first half of the DFT, it follows that a loss function and its gradient with respect to $\hat{D}$ can be defined as:
\begin{subequations}\label{eq:dft_delay_loss}
    \begin{align}
        \mathcal{L} (\hat{D}) &= \sum_{k=0}^{N/2}  |X(k)|^2 \cdot \Gamma(\hat{D}, k),\label{eq:dft_loss}\\
        \quad &\Gamma(\hat{D}, k) = 1 - \cos\left(2\pi k(\hat{D} - D)/{N}\right), \\
        \frac{d \mathcal{L}(\hat{D})}{d {\hat D}} &=  \sum_{k=0}^{N/2} |X(k)|^2 k \sin \left(2\pi k(\hat{D} - D)/N\right) \label{eq:dft_loss_grad}.
    \end{align}
\end{subequations}
The loss function \eqref{eq:dft_loss} and its gradient \eqref{eq:dft_loss_grad} are oscillatory in $\hat{D}$ and the power spectrum $|X(k)|^2$ acts as a frequency-dependent weighting function. For the special case of a spectrally flat input signal,. $|X(k)| = 1$, there is a globally optimal solution at $D = \hat{D}$ but the width of the global minimum is extremely narrow, as shown in Fig.~\ref{fig:dft_delay_colormap} . The implication for gradient--based optimisation is that unless the initial delay estimation is within $\approx \pm 1$ sample of the target delay $D$, the gradients will guide the optimizer to a local minimum and therefore the incorrect solution. The argument to the loss function summation, $\Gamma(\hat{D}, k)$, is visualised in Fig.~\ref{fig:dft_delay_colormap}. Oscillatory local minima are present in both $k$ and $\hat{D}$ directions, with the convex region
% \SB[I would watch out with using this word maybe...it has a pretty specific meaning, and I'm not sure it applies here]
around the global minimum widest for small $k$. We can therefore deduce low-frequency weighting of the loss surface (or using an input signal with a low-passed spectrum) will improve the suitability to gradient-based optimisation.

\begin{figure}[t]
    \centering
    \includegraphics[width=\linewidth]{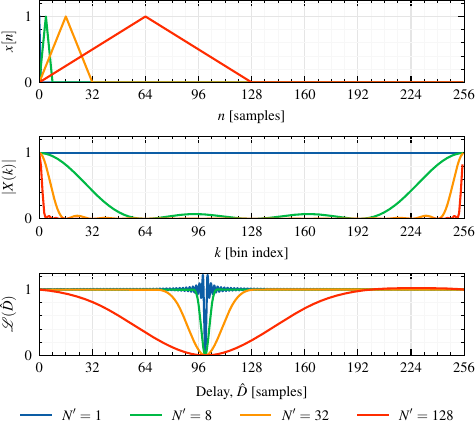}
    \caption{Triangular kernels of length $N'$ (top); their $N=256$ point DFTs (middle); and the corresponding loss surface as a function of $\hat{D}$ for a target delay of $D=100$ samples as in Eq. \eqref{eq:dft_delay_loss} (bottom).}
    \label{fig:fd_mse_delay_windows}
\end{figure}

A simple of choice of a such a signal is a low-pass filter kernel. Fig.~\ref{fig:fd_mse_delay_windows} shows triangular kernels of varying lengths $N'$, their spectra and the resulting loss surface when used as an input signal to the delay optimization system for a DFT length of $N$. The low-frequency weighting of the spectra for $N' > 1$ results in a widening of the convex region around the global minimum to approximately $N'$ samples, making the loss function more suited to gradient-based optimisation.

We can also consider convolving these kernels with signals that have a flat spectrum to give the same resulting loss surface. Due to the linearity of the delay operator, this filtering could be applied before processing with the delay, or afterwards -- in which case the kernel can be viewed as a pre-emphasis filter in the loss function \cite{wright2020_perceptual}. We will consider two signals of this type: the impulse response of a cascade of all-pass filters (AP chirp) \cite{valimaki_spectral_delay}; and a chirp with linearly increasing group delay (Lin chirp) \cite{abel_group_delay2018}. These have a perfectly flat spectrum and are therefore suitable for estimating the frequency response of LTI systems \cite{canfielddafilou:abel:CSNRIR:2018}. Repeated short frames of these signals have also been used in previous works of modelling time-varying modulation effects \cite{Kiiski2016, Wright2021, Carson2023, ycy2024diffapf}.

\subsection{Estimating all-pass coefficients}\label{sec:grad:apf}
\begin{figure}[t]
    \centering
    \includegraphics[width=\linewidth]{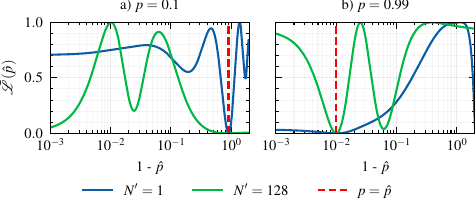}
    \caption{The loss surface as a function of pole location of a cascade of $K = 4$ first order all-pass filer sections for a spectrally flat input signal (length $N'=1$) and a triangular pulse of length $N' = 128$ samples, for a DFT length of $N=256$.}
    \label{fig:apf_loss_tri}
\end{figure}

Suppose now that instead of a linear-phase delay operator, the phase of the input signal is shifted by a cascade of $K$ first-order all-pass filter (APF) sections, as found in phaser effects \cite{DutilleuxDAFx}. Assuming sufficient zero-padding, this can be expressed in the DFT domain as:
\begin{eqnarray}
    Y(k) = X(k) \cdot A_p(k)^K 
\end{eqnarray}
where $A_p(k)$ is the frequency response of the APF section with pole $p$, given by:
\begin{eqnarray}
    A_p(k) = \frac{p - e^{-j2\pi k/N}}{1 - p e^{-j2\pi k/N}}, \quad 0 \leq k \leq N/2.
\end{eqnarray}
The prediction of the pole location $p$ based on observation of $Y(k)$ can formulated as the optimization:
\begin{eqnarray}
    p = \text{arg} \min_{\hat{p}} \mathcal{L}(\hat{p}) 
\end{eqnarray}
where $\hat p$ is the optimization variable and:
\begin{eqnarray}\label{eq:apf_loss}
\begin{aligned}
    \mathcal{L}(\hat{p}) = \sum_{k=0}^{N/2} |X(k)|^2\left|A_p(k)^K - A_{\hat{p}}(k)^K\right|^2 .
    \end{aligned}
\end{eqnarray}
The loss function \eqref{eq:apf_loss} is plotted in Fig.~\ref{fig:apf_loss_tri} for $K=4$ APF sections, two different target pole locations, and two different input signals. Here the loss function is plotted against $1-\hat{p}$, which is approximately proportional to the break frequency of the APF. The loss function shape (not only its shift) depends both on the input signal and the target pole location. In the examples shown, low-frequency weighting does not necessarily improve the loss surface for gradient descent optimization: the local minima are diminished at certain locations but new local minima appear at others. Note however that given the target application of phaser effect modelling (where $K$ is typically limited to $4$ or $6$ APFs), there are fewer ripples overall in the loss function than, for example, when learning a linear-phase delay of several milliseconds in the context of chorus or flanger modelling. Therefore, spectrally flat input signals may indeed be suitable for this application, and indeed previous works indicate that they are \cite{Carson2023, ycy2024diffapf}.

\begin{figure*}[ht]
    \centering
    \includegraphics[width=\textwidth, trim={0 0 6mm 2mm},clip]{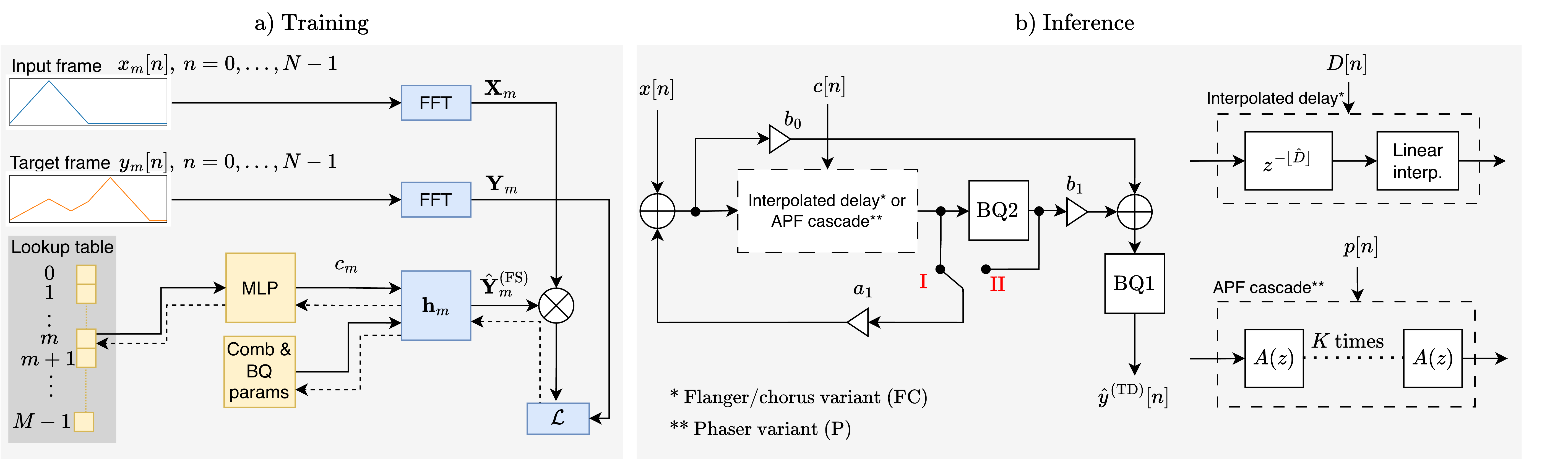}
    \caption{Proposed model structure as it appears during training (a) and at inference (b). Training uses the frequency sampling method over short frames of length $N$ samples. Dashed lines indicate the flow of gradients from the loss function to the learnable modules/parameters (coloured yellow). Inference operates in the time-domain. The switch in position II moves the BQ2 filter within the feedback loop.}
    \label{fig:model}
\end{figure*}

\section{Proposed modelling method}\label{sec:model}
This section outlines the proposed methodology for modelling and sound matching digital and analog time-varying modulation audio effects. 
The model as shown in Fig.~\ref{fig:model} is trained in the time-frequency domain, using a frequency sampling method over short time frames, based on our previous work \cite{Carson2023}. Here however, this framework is used only as a proxy during training, and at inference the learned parameters are employed by a purely time-domain model.

\subsection{Frame-based frequency sampling}\label{sec:model:fs}
The model during training is constrained to process signals for fixed length $L$. An input signal of length $L$ is first windowed into a set of non-overlapping frames $x_m[n], \: n = 0, \dots, N-1$, where $N$ is the frame length and $m$ is the frame index. Here $L$ and $N$ are fixed to be powers of two so that there are exactly $M = L/N$ frames centred at time instants $(m + 0.5) \times N / F_s$ seconds. The FFT of each frame is taken, keeping only the first half of the spectrum to give a set of complex vectors $\textbf{X}_m \in \mathbb{C}^{(N/2+1) \times 1}$. The equivalent STFT for the target audio signal $y[n]$ is taken to give frames $\textbf{Y}_m$, and a linearly spaced frequency vector of length $N/2+1$ samples is initialised, $\textbf{z} = [e^{2\pi j \frac{0}{N}}, \dots, e^{2\pi j \frac{N/2}{N}}]^{T}$. The proposed model predicts a time-dependent frequency response (but time-invariant over the duration of one frame), which is multiplied with the input frame to produce an approximation to the target STFT at frame $m$:
\begin{equation}\label{eq:model:training}
    \hat{\textbf{Y}}^\textrm{(FS)}_m = \textbf{X}_m  \odot \textbf{h}_m
\end{equation}
where $\textbf{h}_m \in \mathbb{C}^{(N/2+1) \times 1}$ is the learned frequency response at frame $m$ and $\odot$ is the Hadamard product. The vectors $\mathbf{h}_m$ are produced by a function with learnable parameters, as detailed in Sec.~\ref{sec:model_freq}, with the time-varying behaviour controlled by an LFO module as detailed in Sec.~\ref{sec:model:lfo}. 

\subsection{LFO generation}\label{sec:model:lfo}
\new[The frame-dependent behaviour of the model is governed by] \new[a learnable control signal $c_m$, $m = 0, \dots, M-1$, that varies at the frame rate $F_f = F_s / N$. In our previous work \cite{Carson2023, ycy2024diffapf}, the control signal was learned by optimising the parameters of an oscillator using the damped sinusoid method  \cite{Hayes2022}. Here, instead, each sample of the control signal is learnable --- the model learns a look-up table (LUT) where each entry is the LFO value for a given frame index. As shown in Fig.~\ref{fig:model}, each sample of the LUT is passed through a small multi-layer perceptron (MLP) neural network. This MLP has a scalar input, a single hidden layer of width 16 with tanh activations, and scalar output with linear activation. The control signal is therefore:]
\new[\begin{equation}
c_m = \mathbf{a}^{T}\tanh(\mathbf{b}\cdot\textrm{LUT}_m + \mathbf{d}) + e
\end{equation}]
\new[where $\{\mathbf{a}, \mathbf{b}, \mathbf{d}\} \in \mathbb{R}^{16\times1}$, $e \in \mathbb{R}$ are the learnable MLP parameters and $\textrm{LUT}_m \in \mathbb{R}$ is the $m$th entry of the learnable look-up table]. \new[The $M$ LUT entries are initialised from a normal distribution $\textrm{LUT}_m \sim \mathcal{N}(\mu, \sigma^2)$ with mean $\mu = 0$ and variance $\sigma^2 = 1/(2\pi)$.  It was found that including the MLP improved robustness to different initialisations rather than setting the control signal as the LUT output directly.]

\new[This proposed method for learning the LFO over-fits the model parameters to the training data over a fixed duration and frame rate. At inference, however, this restriction is lifted by using the learned control signal for wavetable synthesis] \new[\cite{roads_1996}]. \new[The wavetable is constructed by estimating the fundamental frequency of the learned control signal from its magnitude spectrum, truncating to a single period (or multiple thereof) and resampling to the audio sample rate $F_s$. The wavetable is periodically sampled at inference to generate arbitrary length control signals, with the read rate of the wavetable providing a control for the LFO frequency.]

\subsection{Model frequency response}\label{sec:model_freq}
The proposed model at inference consists of a phase shifting filter, either an interpolated delay line or a cascade of first-order APF sections, arranged in the structure shown in Fig.~\ref{fig:model}. During training the model predicts the parameters of a frame-dependent frequency response, given by:
\begin{multline}\label{eq:generic_model}
    \textbf{h}_m  = \textbf{h}_\textrm{\new[BQ1]} \odot \left(b_0 + b_1\textbf{h}_\textrm{\new[BQ2]} \textbf{s}_m\right) \dots \\ \odot \begin{cases}
         \left({1 - a_1 \textbf{s}_m}\right)^{-1} \qquad &\textrm{(I)}\\
            \left(1 - a_1 \textbf{h}_\textrm{\new[BQ2]} \textbf{s}_m \right)^{-1} \qquad &\textrm{(II)}
    \end{cases}
\end{multline}
where $\{b_{0}, b_{1}, a_1\}$ are the learnable comb filter coefficients, $\{\textbf{h}_\textrm{\new[BQ1]}, \textbf{h}_\textrm{\new[BQ2]}\}$ are the frame-invariant frequency responses of learnable biquad \new[(BQ)] filters (defined Sec.~\ref{sec:model:biquad}), and $\mathbf{s}_m$ is the frame-dependent component. \new[We propose two variants of the frame-dependent component]: one for flanger/chorus effects \new[(FC)] and one for phaser effects \new[(P)]. The two cases in \eqref{eq:generic_model} correspond to the switch positions in Fig.~\ref{fig:model}b), with BQ2 either outside (I) or inside (II) the feedback loop. \new[There are therefore 4 distinct configurations of the model which we refer to as FC-I, FC-II, P-I and P-II.]

\textbf{Flanger/chorus variant (FC)}: the frame-dependent component for the flanger/chorus effect is a linear phase delay:
\begin{align}
    \mathbf{s}_m = \textbf{z}^{-d_m}, \quad  d_m = \frac{N}{4}\left(1 - \cos(\pi c_m)\right)
\end{align}
where $c_m$ is the MLP output and the sinusoidal activation maps and bounds the delay between 0 and $N/2$ samples. The lower bound ensures causality, and \new[the upper bound prevents time-aliasing if the input frame is zero-padded by $N/2$ samples --- the input signals used for training were synthesised with this property, as described further in Sec.~\ref{sec:model:training_signals}.]
Model configuration FC-I is used for all flanger/chorus modelling experiments.

\textbf{Phaser variant (P)}: in this case $\mathbf{s}_m$ is the frequency response of a cascade of $K$ \new[identical] APF sections, given by:
\begin{equation}
    \mathbf{s}_m = \left(\left(p_m - \textbf{z}^{-1}\right)\odot \left(1 - p_m \textbf{z}^{-1}\right)^{-1}\right)^{K} 
\end{equation}
The control signal is mapped to the predicted pole location of the APFs with the following activation function:
\begin{eqnarray}\label{eq:pole_activation}
    p_m = \tanh\left(\pi c_m + 0.5\right).
\end{eqnarray}
\new[The function \eqref{eq:pole_activation} bounds the poles of the APF cascade within the unit circle to ensure stability,] and maps the initial distribution towards lower break frequencies. Model configuration \new[P-I] is used by default for the phaser experiments, with a comparison against \new[P-II] given in Sec.~\ref{sec:smallstone}.

\subsection{Biquad filters}\label{sec:model:biquad}
The frequency response of each biquad filter is given by:
\begin{multline}
    \textbf{h}_\textrm{\new[BQ]} = \left({\beta}_0 + {\beta}_1 \textbf{z}^{-1} + {\beta}_2 \textbf{z}^{-2}\right) \dots \\ \odot\left(\alpha_0 + \alpha_1 \textbf{z}^{-1} + \alpha_2 \textbf{z}^{-2}\right)^{-1}.
\end{multline}
Instead of directly optimizing the polynomial coefficients of the filter, we use the state-variable filter parameters to express the coefficients where:
\begin{equation}\label{eq:svf_params}
\begin{aligned}
    \beta_0 &= g^2 m_\textrm{L} + g m_\textrm{B} + m_\textrm{H}, \:\:  &&\alpha_0 = g^2 + 2R + 1\\
    \beta_1 &= 2g^2m_\textrm{L} - 2m_\textrm{H}, \:\: &&\alpha_1 = 2g^2 - 2 \\
    \beta_2 &= g^2m_\textrm{L} - g m_\textrm{B} +  m_\textrm{H}, \:\: &&\alpha_2 = g^2 - 2R + 1.\\
\end{aligned}
\end{equation}
The pole angles depend on $g = \tan(\pi f)$ for normalised frequency $f$ in cycles per sample; $R \geq 0 $ controls the resonance with self-oscillation occurring at $R=0$; and $m_\textrm{L}, m_\textrm{B}, m_\textrm{H}$ are the mixing coefficients between the low-pass, band-pass and high-pass filters. This form been previously favoured over direct form biquad implementations in other differentiable DSP applications \cite{Kuznetsov2020, Lee2022}. Here we use the FLAMO implementation \cite{DalSanto2025_flamo}, which provides non-negativity constraints on $f$ and $R$ to ensure stability:
\begin{equation}
    f :=  \frac{1}{2(1 + e^{-f'})}, \quad R := \log(1 + e^{R'}).
\end{equation}
where $f'$ and $R'$ are the unconstrained parameters learned during training  \cite{DalSanto2025_flamo}. Note that time-aliasing can potentially still occur in our model due to the infinite impulse responses of these filters, but we assume the decay time of the biquad filter will be short relative to the delay or all-pass filters. 
The initialisation of the filter parameters had a significant effect on model robustness with respect to different initial seeds. 
The parameters were initialised as follows: $m_\textrm{\{H, L, B\}} \sim \mathcal{U}(l, u)$ where $\mathcal{U}$ is a uniform distribution with lower and upper bounds $l=0.5$, $u=1.5$; $R'\sim \mathcal{N}(0, 1)$; and $f' \sim \mathcal{N}(-\pi, 1)$. The $-\pi$ here shifts the distribution of $f$ towards low frequencies.

\subsection{Input signals and filters}\label{sec:model:training_signals}
The model as it appears during training is designed for a specific class of input signals that consist of a series of frames of length $N$, such that the input STFT frames $x_m[n]$ (see Sec.~\ref{sec:model:fs}) are identical for all $m$.
Each frame contains a Lin chirp, AP chirp or a triangular kernel (Tri) of length $N'$ samples followed by $N - N'$ zeros. For the chirp inputs, the frequency response of the triangular kernel is considered as an optional pre-emphasis filter in the loss function, denoted $\textbf{h}_{\mathcal{L}}$. A naming convention of \{input signal\} + \{pre-emphasis filter\} is used to distinguish these different combinations.

The numbers $N$ and $N'$ are key hyperparameters. The frame length $N$ controls the trade-off between time and frequency resolution, and, for a given $N$, the non-zero signal length $N'$ controls the trade-off between loss function smoothness, as detailed in Sec.~\ref{sec:grad:delay}, and time-aliasing mitigation both in the model processing and in the target signal measurement. For a target system that implements a delay of $D$ samples, the pulse length must be $N' < N - D$, otherwise the signal in frame $m$ of the input will be delayed into frame $m+1$ of the target. Here we set $N' = N/2$ as we found that this was a suitable trade-off between the competing factors for this application.

\subsection{Loss function, training details and validation}
The loss function is computed directly in the frequency domain as follows:
\begin{equation}\label{eq:complex_esr_preemf}
    \mathcal{L} = \frac{\sum_{m=0}^{M-1} ||\textbf{h}_{\mathcal{L}} \odot \left( \textbf{Y}_m - \hat{\textbf{Y}}^\textrm{(FS)}_m\right)  ||_{2}^{2}}{{\sum_{m=0}^{M-1} ||\textbf{h}_{\mathcal{L}} \odot \textbf{Y}_m  ||_{2}^{2}}}
\end{equation}
where \mbox{$||\cdot||_{2}$} is the L2-norm and $\textbf{h}_{\mathcal{L}}$ 
is the frequency response of the optional pre-emphasis filter.
Models were trained for 15k iterations with an Adam optimizer and learning rate of 1e-3. To test the robustness of training with respect to different initialisations, $R=30$ repetitions of each experiment were performed in parallel using multi-thread processing. Validation is conducted on the time-domain model using guitar and bass inputs of length $L$ samples. The error-to-signal ratio (ESR) and a multi-resolution spectral loss (MRSL) \cite{YamatotoMRSL2020} are computed relative to the target validation output. The 95\% confidence interval (CI) was computed for a given statistic as $\pm 0.95\sigma/\sqrt{R}$ where $\sigma$ is the standard deviation. A sample rate of $F_s = $ \SI{44.1}{\kHz}  and an input signal length of $L=2^{18}$ (\SI{5.94}{\s}) is used throughout. Frame sizes of $N = \{1024, 2048, 4096\}$ (approx. \{23, 46, 93\}\SI{}{\ms}) are considered, giving $M = \{256, 128, 64\}$ frames respectively.

\subsection{Multi-channel extension}
\new[The model can be extended to a $C$-channel implementation where the global output is the sum of the channel outputs. The training and inference outputs are respectively:]
\begin{equation}
        \new[\hat{\textbf{Y}}^\textrm{(FS)}_m = \textbf{X}_m \odot \sum_{\gamma=0}^{C-1} \textbf{h}_{\gamma, m}, \quad
        \hat{y}^\textrm{(TD)}[n]]\ \new[ = \sum_{\gamma=0}^{C-1} \hat{y}^\textrm{(TD)}_\gamma[n]],
\end{equation}
\new[where $\gamma$ is the channel index. This extension is used in the \textit{SV-1} chorus experiments (Sec.~\ref{sec:sv1}). In all other cases $C=1$.]

\section{Toy problem: in-domain parameter recovery}\label{sec:toy}
\begin{figure}[t]
    \centering
    \includegraphics[width=\linewidth]{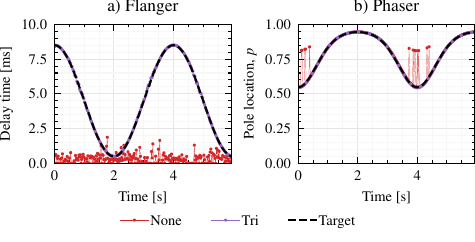}
    \caption{Learned LFOs of the median performing models trained on the AP chirp input with (Tri) and without (None) a pre-emphasis filter in the loss function.}
    \label{fig:digi_lfo}
\end{figure}

As an initial simplified problem, we consider optimizing the proposed model against two specific parametrizations of itself: a flanger configuration and a phaser configuration. The target audio outputs for both effects are generated through the time-domain model (Fig.~\ref{fig:model}b) with the \new[BQ] filters omitted \new[(feedback configurations I and II are equivalent in this case)]. The target flanger effect used the sinusoidal LFO shown in Fig.~\ref{fig:digi_lfo}a). The target phaser effect uses $K=6$ APFs with the shared break-frequency modulated by an LFO that is sinusoidal in the log-frequency space (as is typical \cite{reiss_machpherson_dafx_2014}) as shown in Fig.~\ref{fig:digi_lfo}b). For both effects the feed-forward comb coefficients wet/dry mix is set equal ($b_0 = b_1 = 1$) and the feedback coefficient is set to $a_1 = 0.5$. The starting phases of the LFOs are set to zero for all input signals, including the validation signals, so that there is no phase misalignment when evaluating the model outputs.

In this simplified problem we know there exists a globally optimal solution within the model's parameter space. Note however that the loss metrics can never be exactly zero, because the target signal is generated with an LFO sampled at $F_s$ whereas the trained model learns an approximation at the control rate $F_f$ and then resamples this to $F_s$ at inference with (non-ideal) interpolation. 

\begin{figure}[t]
    \centering
    \includegraphics[width=\linewidth]{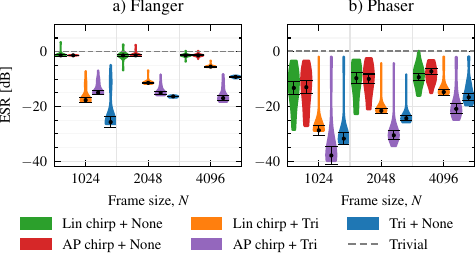}
    \caption{Time-domain ESR for models trained on the digital flanger and phaser effects. Error bars show median and 95\% CI.}
    \label{fig:violins_digi}
\end{figure}

The ESR results of the flanger experiment are shown in Fig.~\ref{fig:violins_digi}a). As a low baseline, the ESR for no model at all is shown with the horizontal grey dashed line -- hereby referred to as the trivial solution. The flanger results show that using the full-band signals (Lin/AP chirp) consistently gives poor results with the median ESR only a few dB below the trivial solution. The results are significantly improved when using the triangular kernel either as the input signal or low-pass pre-emphasis filter, which supports the previous analysis in Sec.~\ref{sec:grad:apf}. The overall best results are achieved using a frame size of $N=1024$ and the Tri + None combination. Examples of the learned LFO signals in the flanger experiment are shown in Fig.~\ref{fig:digi_lfo}a). With no pre-emphasis filter, the model produces a noisy near-zero control signal; with the low-pass pre-emphasis filter the predicted signal accurately follows the target.

The phaser results are shown in Fig.~\ref{fig:violins_digi}b). Again using the triangular kernel either as input or pre-emphasis filter significantly improves the reliability of the results compared to the full-band signals (the 95\% confidence intervals do not overlap), but in this case the Lin/AP chirp models still produce good results for some initial seeds. The AP chirp signal (with no pre-emphasis filter) has been previously used to successfully train phaser models \cite{Carson2023, ycy2024diffapf}, so this result is not surprising. The best overall results were obtained using the AP chirp + Tri combination for $N=1024$. Examples of the learned LFOs for median-performing models are shown in Fig.~\ref{fig:digi_lfo}b). With no pre-emphasis filter, there some discontinuities at the minima of the LFO trajectory; but with the Tri filter these are not present.

\section{BF-2 Flanger}\label{sec:bf-2}
The proposed method was applied to modelling the \textit{Boss BF-2 Flanger} pedal. This is a bucket-brigade delay device \cite{Sangter1969_BBD} with a delay time ranging from \SI{1}{\ms} to \SI{13}{\ms} and an LFO period ranging from \SI{100}{\ms} to \SI{16}{\s} \cite{boss_bf2_manual}. Here we consider two configurations of the pedal with the ``Resonance'' (feedback) knob set to zero (BF-2-A) and to its maximum (BF-2-B). In both cases Rate=0.5 (half-way), Manual=0 and Depth=1, generating an LFO with the largest possible bias and amplitude \cite{boss_bf2_manual}.

We consider only the \{Lin chirp + Tri, AP chirp + Tri, Tri + None\} input/filter combinations as these gave the best results in the previous experiment, and the same choices of $N$ as before. The training and validation signals (each of length $L = 2^{18}$) were concatenated and processed through the pedal to record the target signals. A recording was also taken with the pedal in bypass mode to compensate for any latency \new[caused by the audio interface]. Unlike the simplified experiment, the starting phase of the target LFO cannot be reset so will differ for each training and/or validation signal (unless the LFO period happens to be exactly an integer multiple of $L/F_s$ s). To compensate for this, the \new[fundamental frequency] of control signal $c_m$ is estimated and used to compute its approximate periodic extension to length $2M$. Then the validation loss is computed $M$ times for each starting index $m$, choosing $m$ such that the validation loss is minimised. 

\begin{figure}[t]
    \centering
    \includegraphics[width=\linewidth]{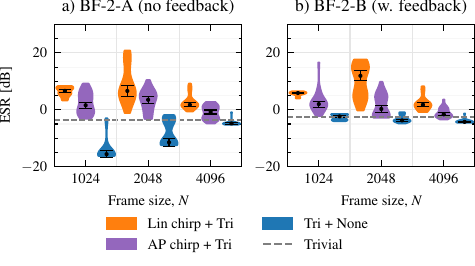}
    \caption{Time-domain validation ESR for models trained on the BF-2 Flanger pedal. Error bars show median and 95\% CI.}
    \label{fig:voilins_bf2}
\end{figure}

\begin{figure}[t]
    \centering
    \includegraphics[width=\linewidth]{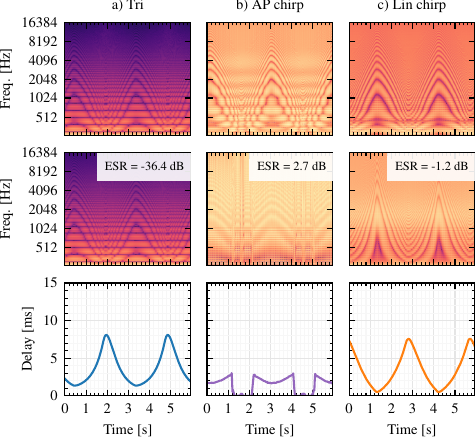}
    \caption{Flanger pedal without feedback (BF-2-A) modelling: mel-spectrograms of target training signal (top), model outputs (mid) and learned LFOs (bottom) for different training inputs (a-c) with $N=1024$. The annotated ESR is the final training loss.}
    \label{fig:bf2-A_specs}
\end{figure}

\new[\textbf{BF-2-A results}:] the ESR results for the pedal with no feedback (BF-2-A) are shown in Fig.~\ref{fig:voilins_bf2}a). Models trained on the Tri input with $N=1024$ give the best  results with the minimum and median ESR both at around \SI{-16}{\dB} (2.5\%). 
\new[The target training signal and model output in this case are shown in Fig.~\ref{fig:bf2-A_specs}a). The spectrograms are visually similar and the waveform ESR (the training loss) is \SI{-36}{\dB}. As shown in the bottom row the learned LFO modulates the delay time between \SI{1}{\ms} and \SI{8}{\ms}.]

The models trained on the chirp inputs (with Tri pre-emphasis filter) produce poor results for all $N$, with the best-case results only slightly better than the trivial threshold. \new[As shown in Fig.~\ref{fig:bf2-A_specs}b-c), models trained on these inputs produce output spectrograms that differ significantly from the target. The discrepancy between these results compared with the the Tri results may potentially be due to some non-linearity in the target pedal --- in this case low-pass filtering at the input vs output would not be equivalent.]

\begin{figure}[t]
    \centering
    \includegraphics[width=\linewidth]{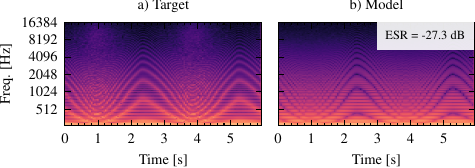}
    \caption{Flanger pedal with feedback (BF-2-B) modelling: mel-spectrograms of a) the target training signal and b) the best model output for the triangular kernel input signal (Tri) with $N=2048$.}
    \label{fig:bf2-B_specs}
\end{figure}

\new[\textbf{BF-2-B results}: ]the modelling of the pedal with feedback (BF-2-B) was overall less successful, as shown by the results in Fig.~\ref{fig:voilins_bf2}b). Here the best-case result (Tri, $N=2048$) is an ESR of \SI{-6}{\dB} (25\%). The target and output spectrograms under these conditions are shown in Fig.~\ref{fig:bf2-B_specs}. The model appears to learn the correct LFO trajectory, but underestimates the amount of feedback: the spectral peaks are much narrower in the target spectrogram than the model output. Further work is required here to resolve this problem. Perhaps a more physically accurate feedback mechanism would give better results, such as including another filter in the feedback path, but initial experiments on this were found to be unsuccessful. 

\section{SV-1 Chorus}\label{sec:sv1}
As a further case study, we consider modelling the \textit{Marshall SV-1 Supervibe} chorus pedal. It features four knobs: ``Speed'', controlling the LFO frequency; ``Depth'' controlling the input/delayed signal mix (note that this is different to the \textit{BF-2} control); a ``Wave'' control that blends a fixed LFO with the variable speed LFO controlled by the ``Speed'' knob; and a ``Filter'' knob which is a tone control on the delayed signal \cite{sv1_manual}. Here we consider three configurations: Depth=0.5, Wave=0 (SV-1-A); Depth=0.5, Wave=1 (SV-1-B); and Depth=1, Wave=1 (SV-1-C). In all three cases Rate=0.5 and Filter=1.0.

\begin{figure}[t]
    \centering
    \includegraphics[width=\linewidth]{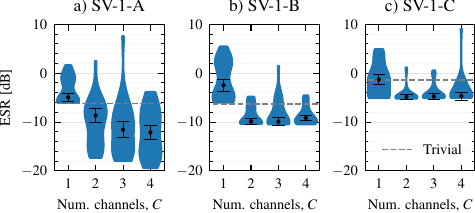}
    \caption{Time-domain validation ESR results for models trained on three configurations of the \textit{SV-1} chorus pedal for different number of model channels. Error bars show median and 95\% CI.}
    \label{fig:violins_sv1}
\end{figure}

\begin{figure}[t]
    \centering
    \includegraphics[width=\linewidth]{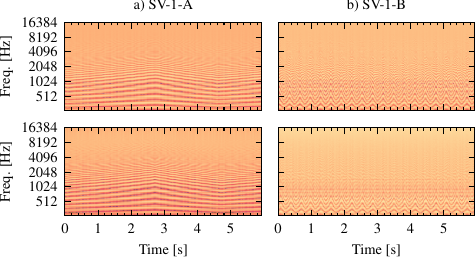}
    \caption{Target (top) and model (bottom) output mel-spectrograms for the \textit{SV-1} chorus pedal with the ``Wave'' control at minimum (SV-1-A) and maximum (SV-1-B).}
    \label{fig:sv-1_specs}
\end{figure}

In this case study the effect of the multi-channel variant of the model is considered. The input signal was fixed to the triangular kernel (Tri) with $N=1024$ as this gave the best results in the \textit{BF-2} experiment. The results are shown in Fig.~\ref{fig:violins_sv1}. 
The SV-1-A and SV-1-B results (Depth=0.5), show that a minimum of $C=2$ model channels are required to produce results that exceed the trivial threshold, but beyond this there is only marginal improvement. The SV-1-C (Depth=1) results show that models with $C=1$ can achieve as good results as $C=2$ but less reliably---the median ESR is at the trivial threshold for the mono variant. These result may suggest that the model uses one channel for the dry path, and the other for the delay-modulated (wet) signal, but investigation of the individual channels found that this is not necessarily the case: there are multiple combinations of parameter configurations that produce similar outputs. Target and model outputs are shown in Fig.~\ref{fig:sv-1_specs} for the SV-1-A and SV-1-B configurations. With the ``Wave'' control at zero (Fig.~\ref{fig:sv-1_specs}a) the pedal acts as a subtly varying comb filter; and at its maximum (Fig.~\ref{fig:sv-1_specs}b) higher frequency oscillations are introduced to produce a chorus effect.

\section{Small Stone Phaser}\label{sec:smallstone}
 As a final case study the phaser variant of the model is applied to the \textit{EHX Small Stone} pedal. This pedal has been modelled in previous work using a similar approach to the one presented here \cite{Carson2023, ycy2024diffapf}. Here we develop this work by investigating the effect of different input signals and pre-emphasis filters on the robustness of training. Two configurations are considered with the ``Rate’’ control set to half-way and the ``Color’’ (feedback) switch off/on (SS-A/SS-B). The pedal has $K=4$ APF sections and the model is manually configured accordingly.

\begin{figure}[t]
    \centering
    \includegraphics[width=\linewidth]{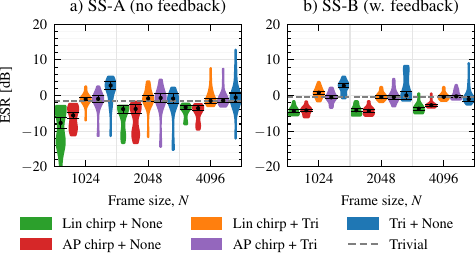}
    \caption{Time-domain validation ESR for models trained on the \textit{Smallstone} phaser pedal. Error bars show median and 95\% CI.}
    \label{fig:violins_ss}
\end{figure}

\new[The validation ESR results (Fig.~\ref{fig:violins_ss}) show that the spectrally flat signals (AP chirp + None, Lin chirp + None) give more consistent results than the low-frequency weighted signals, especially when modelling the pedal with feedback enabled (SS-B). Note that this is the opposite result to the flanger modelling experiment (Sec.~\ref{sec:bf-2}) where low-pass filtering proved essential for accurate results. ]

\begin{figure}[t]
    \centering
    \includegraphics[width=\linewidth]{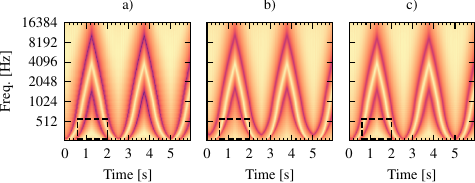}
    \caption{\textit{Smallstone} modelling: target output (a) and model outputs without (b) and with (c) filtering in the feedback loop. \new[Dashed boxes highlight notable differences in the spectrograms.]}
    \label{fig:ss_specs}
\end{figure}

As a final experiment, a comparison of the two feedback configurations of the model in \eqref{eq:generic_model} is explored, with results shown in Table~\ref{tab:fb_compare}. Of particular interest are the SS-B results, as these are for the pedal with ``Color'' (feedback) activated. The median ESR is very similar for both feedback configurations, but the confidence interval is larger for models with filtering in the feedback loop (P-II) --- thus this method is less reliable for producing consistent results. In the best-case scenario, configuration PB-II gives an ESR improvement of \SI{2.2}{\dB}, and while this is small informal critical listening found that the difference in sound outputs between the models was audible. Spectrograms are shown in Fig.~\ref{fig:ss_specs} to illustrate this: the difference is subtle, but the high energy regions in the lower frequencies are better emulated by model P-II than model P-I.

\begin{table}[t!!]
    \centering
    \resizebox{\linewidth}{!}{
    \begin{tabular}{cccc}
    \hline
    Effect & \new[Model config.] & Median ESR [dB] & Best ESR [dB] \\
    \hline
    \multirow{2}{*}{SS-A} & \new[P-I] & -2.6 $\pm$ 1.0 & -14.4 \\ 
    & \new[P-II] & -3.0 $\pm$ 7.3 & -14.4 \\ 
    \hline
    \multirow{2}{*}{SS-B}   & \new[P-I] & -1.6 $\pm$ 0.6 & -5.7 \\ 
    & \new[P-II] & -1.5 $\pm$ 9.5 & -7.9 \\ 
    \hline
    \end{tabular}}
    \caption{Comparison of feedback configurations for modelling the \textit{Smallstone} phaser, using the AP chirp input with $N=2048$ over 30 initial seeds.}
    \label{tab:fb_compare}
\end{table}

\begin{table*}[ht]
    \centering
    \resizebox{\linewidth}{!}{
    \begin{tabular}{cccccccccccc}
    \hline
        Effect & \new[Model config.] &Training input & $N$ & $C$ & Seed & ESR [dB] & MRSL &  Ref. & Model & Anchor \\
    \hline
BF-2-A & \new[FC-I] &Tri & 2048 & 1 & 3 & -12.39 & 0.67 & 98.8 $\pm$ 1.7$^*$ & 92.9 $\pm$ 4.6$^*$ & 1.9 $\pm$ 1.7\\ 
    BF-2-B & \new[FC-I] & Tri & 2048 & 1 & 0 & -4.53 & 1.03 & 100 $\pm$ 0.0 & 42.0 $\pm$ 6.0 & 0.7 $\pm$ 0.8\\ 
    SV-1-B & \new[FC-I] & Tri & 1024 & 2 & 25 & -10.27 & 0.68 & 98.4 $\pm$ 2.2$^*$ & 95.0 $\pm$ 4.2$^*$ & 1.3 $\pm$ 1.1\\
    SV-1-C & \new[FC-I] & Tri & 1024 & 2 & 25 & -5.10 & 0.88 & 100 $\pm$ 0.0 & 89.5 $\pm$ 4.9 & 1.3 $\pm$ 1.4\\ 
    SS-A & \new[P-II] & AP chirp & 2048 & 1 & 8 & -13.57 & 0.45 & 100 $\pm$ 0.0$^*$ & 98.9 $\pm$ 1.7$^*$ & 6.4 $\pm$ 7.8\\ 
    SS-B & \new[P-II]& AP chirp & 2048 & 1 & 9 & -7.84 & 0.76 & 98.8 $\pm$ 1.8$^*$ & 98.3 $\pm$ 1.7$^*$ & 1.0 $\pm$ 1.4\\ 
    \hline
    \end{tabular}
    }
    \caption{Training configurations, objective metrics and mean similarity ratings $\pm$ 95\% confidence intervals (CIs) for the models used in the listening test. $^*$ indicates no statistical significance (overlapping CIs) between Model and Reference ratings. }
    \label{tab:results}
\end{table*}
\section{Perceptual evaluation}\label{sec:mushra}
The perceived quality of six selected models compared to the analog reference devices was investigated with a MUSHRA-style listening test \cite{MUSHRA_ITU-R}. The test consisted of 18 trials using three different guitar excerpts as input signals per effect. Participants were presented with the output from the analog reference, and asked to rate the model output, hidden reference and an anchor signal based on the perceived similarity to the reference. The anchor was the uneffected input signal ---  we acknowledge that this is a very basic baseline but wanted to keep test complexity low to avoid listener fatigue. Twelve participants took the test, with 10 remaining after standard post-screening \cite{MUSHRA_ITU-R}.

The results are shown in Table~\ref{tab:results}. For the BF-2-A, SV-1-B, SS-A and SS-B models, there is no statistical significance between the Reference and Model ratings suggesting that the models are indistinguishable from the analog references. The SV-1-C model (the vibrato effect) was rated lower but still within the MUSHRA Excellent band \cite{MUSHRA_ITU-R}. The model trained on the flanger with feedback (BF-2-B) gives the worst results, with the mean rating in the Fair band, and this is consistent with the objective evaluation.

\section{Conclusion}\label{sec:conclusion}
This work investigated the optimisation of phase shifting modulation effects including flangers, chorus and phasers to match digital and analog reference systems. For learning delay times (relevant for chorus and flanger effects), low-frequency weighting of the loss function using either a low-pass kernel input signal or pre-emphasis filter proved essential for model convergence. The results for cascaded all-pass filter coefficient estimation (relevant to phaser effects) were less conclusive, but full-band optimisation yielded good results on an analog phaser modelling experiment. A perceptual evaluation showed that the proposed model gave an excellent perceived similarity to analog references in most cases, but devices which exhibit long delay times with feedback proved more challenging to model.

\bibliographystyle{IEEEtran}
\bibliography{refs}

@book{roads_1996,
	Address = {Cambridge, MA},
	Author = {Roads, C},
	Edition = 1,
	Publisher = {MIT Press},
	Title = {The Computer Music Tutorial},
	Year = {1996}}

@InProceedings{Hayes2022,
  author={Hayes, B. and Saitis, C. and Fazekas, G.},
  booktitle={IEEE ICASSP}, 
  title={Sinusoidal Frequency Estimation by Gradient Descent}, 
  year={2023},
  month={June},
  volume={},
  number={},
  address={Rhodes, Greece}}

@article{Wright2021,
   author = {A. Wright and V. Välimäki},
   issue = {7-8},
   journal = {J. Audio Eng. Soc.},
   number = {7},
   pages = {517-529},
   publisher = {Audio Engineering Society},
   title = {Neural Modeling of Phaser and Flanging Effects},
   volume = {69},
   year = {2021},
}

@article{Wright2020,
   author = {A. Wright and E.-P. Damskägg and L. Juvela and V. Välimäki},
   issn = {20763417},
   issue = {3},
   journal = {Appl. Sci.},
   number = {2},
   publisher = {MDPI AG},
   title = {Real-time guitar amplifier emulation with deep learning},
   volume = {10},
   year = {2020},
   artcile-number = {766},
   doi = {10.3390/app10030766},
}

@InProceedings{Engel2020,
   author = {J. Engel and L. Hantrakul and C. Gu and A. Roberts},
   booktitle = {Int. Conf. Learning Repr.},
   title = {{DDSP}: Differentiable digital signal processing},
   year = {2020},
}

@InProceedings{Kuznetsov2020,
   author = {B. Kuznetsov and J. Parker and F. Esqueda},
   booktitle = {23rd Int. Conf. Digital Audio Effects (DAFx20)},
   title = {Differentiable {IIR} filters for machine learning applications},
   year = {2020},
   month = {Sept.},
   address = {Vienna, Austria}
}

@article{bartlett1970a,			  		  
author={B. Bartlett},					  
journal={J. Audio Eng. Soc.}, 					  
title={A scientific explanation of phasing (flanging)}, 					  
year={1970},					  
volume={18},					 
number={6},					  
pages={674--675},					  
doi={},}

@book{PASPWEB2010,
	AUTHOR = "J. O. {Smith III}",
	TITLE = "Physical Audio Signal Processing",
	PUBLISHER = "\url{http://ccrma.stanford.edu/~jos/pasp/}",
	YEAR = "accessed 10/3/26",
    NOTE = "online book, 2010 edition"
}

@InProceedings{Eichas2014,
   author = {F. Eichas and M. Fink and M. Holters and U. Zölzer},
   city = {Erlangen, Germany, },
   booktitle = {17th Int. Conf. Digital Audio Effects (DAFx14)},
   month = {Sept.},
   address = {Erlangen, Germany},
   title = {Physical modeling of the {MXR} {P}hase 90 guitar effect pedal},
   year = {2014},
}

@InProceedings{Kiiski2016,
   booktitle = {19th Int. Conf. Digital Audio Effects (DAFx16)},
   address = {Brno, Czech Republic},
   month = {Sept.},
   year = {2016},
   author = {R. Kiiski and F. Esqueda and V. Välimäki},
   title = {Time-variant gray-box modeling of a phaser pedal},
}

@article{bode1984history,
		title = {History of Electronic Sound Modification},
		author = {H. Bode},
		journal = {J. Audio Eng. Soc.},
		volume = {32},
		number = {10},
		pages = {730--739},
		year = {1984}
		}

@incollection{DutilleuxDAFx,
author = {Dutilleux, P. and Holters, M. and Disch, S. and Zölzer, U.},
publisher = {John Wiley \& Sons, Ltd},
isbn = {9781119991298},
title = {Filters and Delays},
booktitle = {DAFX: Digital Audio Effects},
chapter = {2},
pages = {47-81},
year = {2011},
}

@incollection{DAFXVAchapter,
pages = {473--522},
publisher = {John Wiley \& Sons, Ltd},
booktitle = {DAFX: Digital Audio Effects},
year = {2011},
title = {Virtual Analog Effects},
copyright = {2011 Wiley},
language = {eng},
author = {Välimäki, V. and Bilbao, S. and Smith, J. O and Abel, J. S and Pakarinen, J. and Berners, D.}}

@ARTICLE{Laakso1996,
  author={Laakso, T. I. and Välimäki, V. and Karjalainen, M. and Laine, U. K.},
  journal={IEEE Signal Process. Mag.}, 
  title={Splitting the unit delay}, 
  year={1996},
  month={Jan.},
  volume={13},
  number={1},
  pages={30-60}}

@inproceedings{Carson2023,
title = "Differentiable grey-box modelling of phaser effects using frame-based spectral processing",
author = "Alistair Carson and Simon King and {Valentini Botinhao}, Cassia and Stefan Bilbao",
year = "2023",
month = sep,
day = "4",
language = "English",
booktitle = "26th Int. Conf. on Digital Audio Effects",
}

@inproceedings{wright2020_perceptual,
author = {Wright, Alec and Välimäki, Vesa},
year = {2020},
month = {May},
pages = {251-255},
title = {Perceptual loss function for neural modeling of audio systems},
booktitle={IEEE Int. Conf. Acoust. Speech Signal Process. (ICASSP)}
}

@INPROCEEDINGS{YamatotoMRSL2020,
  author={Yamamoto, Ryuichi and Song, Eunwoo and Kim, Jae-Min},
  booktitle={IEEE ICASSP.}, 
  title={Parallel {WaveGAN}: A Fast Waveform Generation Model Based on Generative Adversarial Networks with Multi-Resolution Spectrogram}, 
  address = {Barcelona, Spain},
  year={2020},
  volume={},
  number={},
  doi={10.1109/ICASSP40776.2020.9053795}}

@article{valimaki_spectral_delay,
author = {Välimäki, Vesa and Abel, Jonathan and Smith, Julius},
year = {2009},
month = {07},
pages = {521-531},
title = {Spectral Delay Filters},
volume = {57},
journal = {J. Audio Eng. Soc.}
}

@inproceedings{canfielddafilou:abel:CSNRIR:2018,
    author = {Elliot Canfield-Dafilou and Jonathan Abel},
    title = {An Allpass Chirp for Constant Signal-to-Noise Ratio Impulse Response Measurement},
    booktitle = {144th Audio Engineering Society Convention},
    year = {2018}}

@inproceedings{Holters2018_BBD,
   author = {Martin Holters and Julian D Parker},
   city = {Aviero, Portugal},
   booktitle = {21st Int. Conf. Digital Audio Effects (DAFx-18)},
   title = {A Combined Model for a Bucket Brigade Device and its Input and Output Filters},
   year = {2018}
}

@ARTICLE{Sangter1969_BBD,
  author={Sangster, F.L.J. and Teer, K.},
  journal={IEEE Journal of Solid-State Circuits}, 
  title={Bucket-brigade electronics: new possibilities for delay, time-axis conversion, and scanning}, 
  year={1969},
  volume={4},
  number={3},
  pages={131-136},
  doi={10.1109/JSSC.1969.1049975}}

@inproceedings{DalSanto2025_flamo,
   author = {Gloria Dal Santo and Gian Marco De Bortoli and Karolina Prawda and Sebastian J. Schlecht and Vesa Välimäki},
   isbn = {9798350368741},
   issn = {15206149},
   booktitle = {IEEE ICASSP},
   title = {{FLAMO}: An Open-Source Library for Frequency-Domain Differentiable Audio Processing},
   year = {2025}
}

@manual{boss_bf2_manual,
  title        = {Boss {BF-2} Flanger instructions},
  organization = {Roland},
  address      = {Japan},
  month =       {July},
  year         = {1985}
}

@manual{sv1_manual,
  title        = {{SV-1} Supervibe chorus instructions},
  organization = {Marshall},
}

@inproceedings{ycy2024diffapf,
  title={Differentiable All-pole Filters for Time-varying Audio Systems},
  author={Chin-Yun Yu and Christopher Mitcheltree and Alistair Carson and Stefan Bilbao and Joshua D. Reiss and György Fazekas},
  booktitle={27th Int. Conf. Digital Audio Effects (DAFx24)},
  year={2024}
}

@book{reiss_machpherson_dafx_2014,
author = {Reiss, Joshua D. and McPherson, Andrew},
address = {Boca Raton, FL},
booktitle = {Audio effects : theory, implementation and application},
edition = {1st edition},
isbn = {9781040061589},
language = {eng},
publisher = {CRC Press, an imprint of Taylor and Francis},
title = {Audio effects : theory, implementation and application },
year = {2014},
}

@techReport{JuliusOSmith1982,
   author = {Julius O. Smith},
   title = {An allpass approach to digital phasing and flanging},
   year = {1982}
}

@inproceedings{Huovilainen_2005,
   author = {Antti Huovilainen},
   city = {Madrid, Spain},
   booktitle = {3rd Int. Conf. Digital Audio Effects (DAFx05)},
   title = {Enhanced digital models for analog modulation effects},
    month = {9},
year = {2005}
}

@inproceedings{Lee2024,
   author = {Gyubin Lee and Hounsu Kim and Junwon Lee and Juhan Nam},
   booktitle = {27th Int. Conf. Digital Audio Effects (DAFx24)},
   month = {6},
   title = {CONMOD: Controllable Neural Frame-based Modulation Effects},
year = {2024}
}

@book{oppenheim189_dsp,
author = {Oppenheim, Alan V. and Schafer, Ronald W.},
booktitle = {Discrete-time signal processing},
isbn = {0132167719},
language = {eng},
publisher = {Prentice-Hall International},
title = {Discrete-time signal processing },
year = {1989},
}

@book{MUSHRA_ITU-R,
    author = {},
    title = {Method for the subjective assessment
of intermediate quality level
of audio systems},
    address = {ITU-R recommendation BS.1534-3}
}

@ARTICLE{Lee2022,
  author={Lee, Sungho and Choi, Hyeong-Seok and Lee, Kyogu},
  journal={IEEE Trans. on Audio, Speech, and Language Processing}, 
  title={Differentiable Artificial Reverberation}, 
  year={2022},
  volume={30},
  number={},
  pages={2541-2556},
  doi={10.1109/TASLP.2022.3193298}}

@inproceedings{abel_group_delay2018,
    author = {Elliot K. Canfield-Dafilou and Jonathan S. Abel} ,
    title = {Group delay-based allpass filters for abstract sound synthesis and audio effects processing},
    booktitle = {21st Int. Conf. Digital Audio Effects (DAFx18)},
    year = {2018},
    address = {Aveiro, Portugal},
    month = {9}
}

@ARTICLE{hayes_review_2023,
  AUTHOR={Hayes, Ben  and Shier, Jordie  and Fazekas, György  and McPherson, Andrew  and Saitis, Charalampos },   
TITLE={A review of differentiable digital signal processing for music and speech synthesis},   
JOURNAL={Frontiers in Signal Processing},   
VOLUME={3},
YEAR={2024},
DOI={10.3389/frsip.2023.1284100},
ISSN={2673-8198}}

@inproceedings{Mitcheltree2023,
   author = {Christopher Mitcheltree and Christian J. Steinmetz and Marco Comunità and Joshua D. Reiss},
   booktitle = {26th Int. Conf. Digital Audio Effects (DAFx23)},
   month = {5},
   title = {Modulation Extraction for LFO-driven Audio Effects},
   year = {2023}
}

@inproceedings{Mitcheltree2025,
   author = {Christopher Mitcheltree and Hao Hao Tan and Joshua D. Reiss},
   booktitle = {IEEE WASPAA},
   title = {Modulation Discovery with Differentiable Digital Signal Processing},
   year = {2025}
}

@inproceedings{Giampiccolo2024_phase90,
    author = {Riccardo Giampiccolo and Samuele Del Moro and Claudio Eutizi and Mattia Massimi, Oliviero Massi and Alberto Bernardini},
    title = {Wave digital model of the {MXR} {Phase} 90 based on a time-varying
Resistor approximation of {JFET} elements
},
    booktitle = {27th Inf. Conf. on Digital Audio Effects (DAFx24)},
    year = {2024},
    address = {Guildford, UK},
    month={9}
}

@Article{Ramirez2020_blackbox,
AUTHOR = {Martínez Ramírez, Marco A. and Benetos, Emmanouil and Reiss, Joshua D.},
TITLE = {Deep Learning for Black-Box Modeling of Audio Effects},
JOURNAL = {Applied Sciences},
VOLUME = {10},
YEAR = {2020},
NUMBER = {2},
ARTICLE-NUMBER = {638},
ISSN = {2076-3417}}

\end{document}